**Italy 100% Renewable: A Suitable Energy Transition Roadmap**

Francesco Meneguzzo,[a*] Rosaria Ciriminna,[b] Lorenzo Albanese,[b] Mario Pagliaro[b*]

[a]*Istituto di Biometeorologia, CNR, via G. Caproni 8, 45045 Firenze, Italy;* [b]*Istituto per lo Studio dei Materiali Nanostrutturati, CNR, via Ugo La Malfa 153, 90146 Palermo, Italy*

**Abstract**

We outline a realistic energy transition roadmap for Italy, in which the whole energy demand is met by electricity generated by low cost renewable energy technologies, namely solar photovoltaic, wind and hydroelectric power. We assess the amount of extra power and storage capacity to be installed along with costs, return on investment and payback time. Based on cost, renewable nature and scalability, storage in energy dense polysaccharides enzymatically synthesized from carbon dioxide, water and surplus electricity is proposed to meet the significant storage requirements.

*\*Corresponding authors*

Dr. F. Meneguzzo
Istituto di Biometeorologia, CNR
via G. Caproni 8
45045 Firenze (Italy)
E-mail: francesco.meneguzzo@cnr.it

Dr. M. Pagliaro
Istituto per lo Studio dei Materiali Nanostrutturati, CNR
via U. La Malfa 153
90146 Palermo PA (Italy)
E-mail: mario.pagliaro@cnr.it

# 1 Introduction

The dramatic increase of renewable energy generation across the world which occurred in the last decade has been so significant and rapid that, for short periods of time, the electricity demand of whole industrial countries has been temporarily met by means of renewable energy, notably wind, hydro and photovoltaic power. For instance, as of May 15, 2016 (a Sunday) at 2 pm local time when demand in Germany was 45.8 GW renewable energies covered around 82% of power demand.[1] Since about 2008, scholars,[2] energy analysts[3] and environmental groups[4] started to investigate the feasibility of a full transition from fossil to renewable energy.

The most radical energy transition scenario was likely proposed by Jacobson and Delucchi in 2009 by investigating the feasibility of transitioning entire countries by 2030 to energy systems powered exclusively by wind, water and sunlight.[5] The team calculated the number of onshore and offshore wind, photovoltaic (PV, on rooftops and utility scale), concentrated solar power (CSP), geothermal power, tidal and wave power, and existing hydroelectric generators needed to power each country based on the 2050 power demand after all energy end uses (namely electricity, heating and cooling, and transport) have been electrified. In detail, to supply the world with 100% renewable energy (electricity and electrolytic hydrogen) would require 3.8 million 5-MW wind turbines, 40,000 300-MW central solar plants, 40,000 300-MW solar PV plants, 1.7 billion 3-kW rooftop PV installations, 5,350 100-MW geothermal plants, 270 new 1.3-GW hydro stations, 720,000 0.75-MW wave devices, and 490,000 1-MW tidal turbines.

Smil has remarked the scope of such a transition noting that, according to this scenario, in just fifteen years the overall installed capacities would have to increase 30-fold for wind, 100-fold for geothermal power and 500-fold for tidal power, with 40,000 new large (300 MW) PV plants and nearly 50,000 new CSP plants, as well as more than 700,000 wave-conversion projects.[6]

For decades, critics of renewable energy lamented their unsustainably high cost.[7] Since the early 2000s, however, PV[8] and wind[9] energy are experiencing a largely unexpected global growth (178 GW of solar PV and 433 GW of wind power installed across the world by 2015), which has led the cost of clean electricity to such low levels to go below that of electricity obtained by burning coal ($2.3c/kWh in a in an auction by Abu Dhabi for a 350MW solar plant as of September 2016).[10] A similar trend has occurred for the commonly neglected solar thermal technology, with solar heating contribution in meeting global energy demand second only to wind power among renewable energy sources (341 TWh of energy supplied in 2014 mainly as hot water *vs* 200 TWh of electricity supplied by PV modules across the world).[11]

Critics of renewable energy today preferably advocate the "much lower quality"[12] of renewable energy whose intermittent nature would pose insurmountable economic problems to the grid reliability. For example, while convening that it would be technically possible to meet total electricity demand from renewable energy sources, Trainer argues that this would be unaffordable due to the amount of redundant plants needed to cope with intermittency.[13] However, wind and solar energy complement each other on intraday and seasonal scales (continental wind energy tends to peak at night, solar during the day, while the wind blows during winter and stormy days when solar modules produce little electricity). The higher the number of wind turbines and solar panels connected to the grid, and the more geographically distributed across a territory, the less volatile is the combined output of all individual generators, smoothing out renewable energy fluctuations on a second-by-second basis.[14] Neither Germany, Spain nor Italy, for example, faced particular grid problems when a large amount of solar PV and wind power was installed in the 2005-2015 decade.

In this study, we evaluate the full transition of Italy to renewable energy by the year 2050. We adopt a realistic approach, which starts from the very awareness that production, installation, and maintenance of wind turbines and PV modules "remains critically dependent on specific fossil energies";[15] while so far the fastest historical sector-specific energy transitions observed was 30 years, though energy transitions involving all sectors have taken much longer.[16] The outcomes of this work will be useful to policy makers and energy stakeholders called by energy and environmental urgencies to accelerate the transition to renewable energy.

## 2 Current scenario

The set of data and their respective sources listed in Table 1 include the relevant information about the Italian energy system.

**Table 1**. Italy's energy datasets and sources

| Dataset | Period and frequency | Unit | Source |
|---|---|---|---|
| Energy consumption by source[a] | 1965-2015 Annual | MTOE | BP[17] |
| Power capacity by source (excl. hydroelectric) | | | |
| Hydroelectric capacity by type | 1999-2015 Annual | MW | GSE[18] |
| Electricity consumption | 2014-2015 Hourly | TWh | Terna[19] |
| Intermittent generation from RES, by source | 2014-2015 Hourly | TWh | Terna[20] |
| Oil consumption for transportation | 2014-2015 Monthly | MTOE | MISE[21] |

[a]The electricity consumption in units of TWh is converted into primary energy in units of MTOE after accounting for an average thermoelectric efficiency factor equal to 0.38, as well as for the average oil energy content equal to 11.6 kWh/kg.[1]

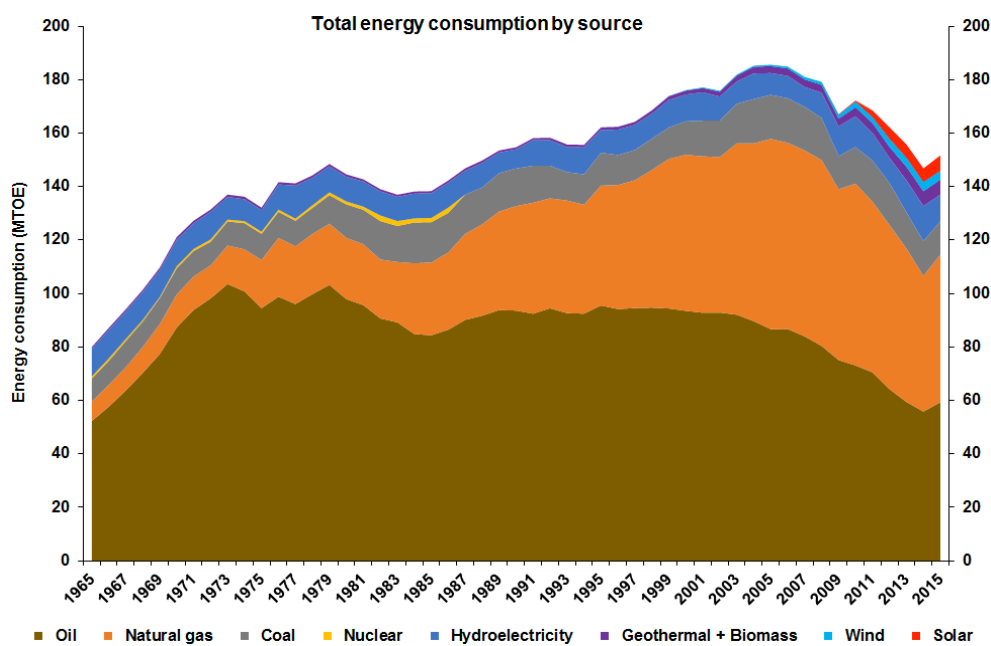

**Figure 1**. Total energy consumption by source in Italy, during 1960-2015.

Figure 1 shows that the total energy consumption (TEC) in Italy has dropped as much as 18% in the period 2005-2015 (from 185.6 to 151.7 MTOE). Followed by coal, reduction in oil and natural gas use was particularly pronounced. Such reduction was only partially compensated by a relevant increase of renewable energy sources (RES), mainly wind and solar PV. The share of RES (Figure 2) over TEC has quickly risen from a fairly stable 6% until 2007-2008 to over 18% in 2014, while dropping to slightly above 16% in 2015 due to climate-dependent reduction of hydroelectric output.

Remarkably, these figures match the binding requirements laid down by the European Directive 2009/28/EC of the European Parliament and of the Council of 23 April 2009 on the promotion of the use of energy from renewable sources,[22] according to which in Italy 17% of gross end energy consumption in 2020 should be met by renewable energy sources.

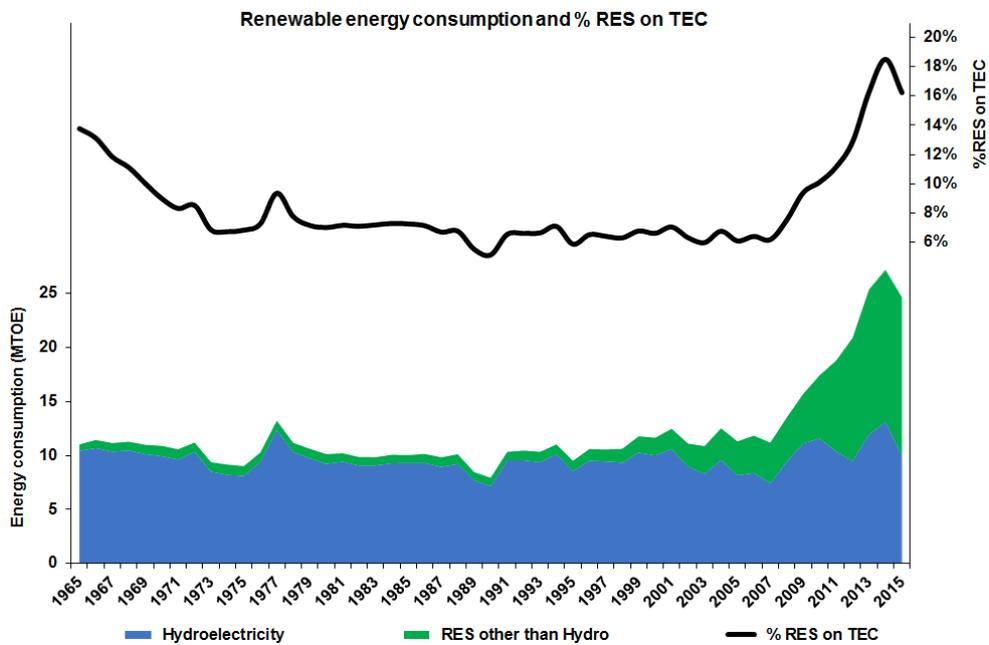

**Figure 2**. Renewable energy generation partitioned into hydroelectricity and other sources, along with its share of total energy consumption, in Italy, during 1960-2015.

In practice, the whole increase of RES generation was produced by sources other than hydroelectric power. Since 1999, the wind source has steadily increased, totaling as much as 9,000 MW (Figure 3), while the solar PV source has steeply risen from practically zero MW in 2007 to 19,000 MW in 2015, with most of the increase (+16,000

MW) during 2008-2012, and more specifically in 2011 and 2012, when Italy's photovoltaic capacity grew by about 15 GW and onshore wind by about 4 GW.[6] In the same period, the capacity of hydroelectric power increased by 2,000 MW; and that of geothermal power by 150 MW only. At the end of 2013, Italy had an installed 18.5 GW of PV (+18.500 MW from 2005) and 8.4 GW of wind (+7.000 MW from 2005) power, with most renewable energy plants in the South, far from large electricity consumers.[23] Concomitantly, the electricity demand declined from the 2007 peak (340 TWh) to 318 TWh in 2013, *i.e.* even lower than the 2003 level (321 TWh).

Together, the boom in RES generation and the decline in demand brought the equivalent hours of natural gas power plants in operation from 5,100 of 2006 to 2,100 as of 2013 (-60%).[23] In 2014 Italy's incumbent generator announced the decommissioning of 13 GW of thermal capacity within five years (of which it has decommissioned 8 GW to date, starting from the coal-fired station near Venice).

To cope with the surge in RES generation and decline in demand, the State-owned grid company Terna invested about 5 billion € in upgrading the grid by building 2.000 km of new lines and 70 new substations, and implemented numerous innovative solutions (management of active and reactive power through new transformers and capacitors, optimization of the existing lines and accurate prediction of RES generation). As a result, curtailment of renewable energy plants went from 9.7% in 2009 to 0.4% in 2013.[23]

Figure 3 shows that the whole increase of hydroelectric capacity in 1999-2015 was due to new off-river hydroelectric plants, which are unable to store electricity generated by intermittent sources, while the capacity of reservoir installations (more than 400 hours of storage) and basins (2-400 hours) has been fairly stationary.

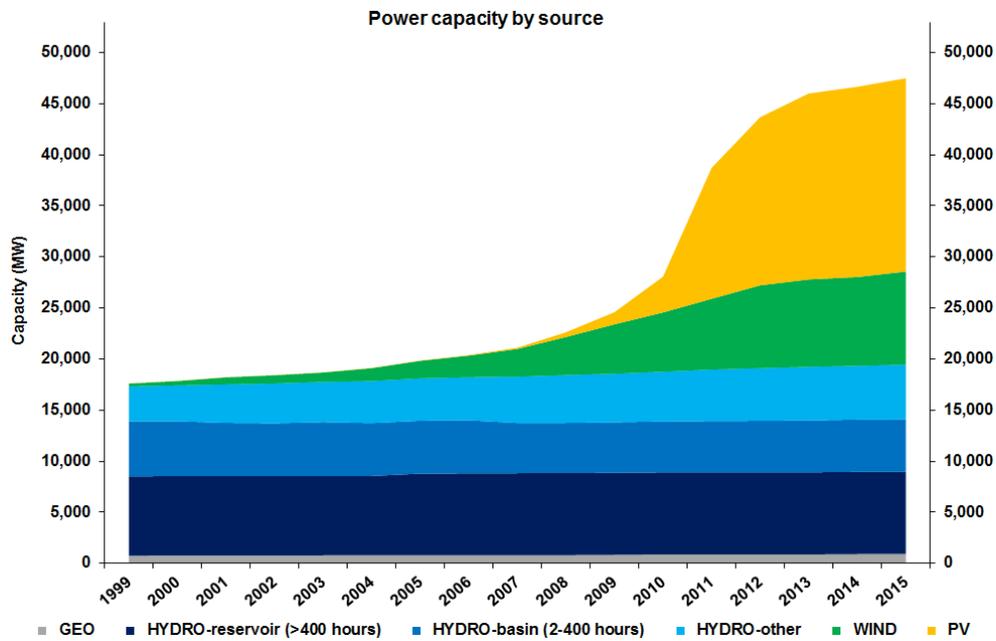

**Figure 3**. Power capacity of renewable energy sources in Italy, during 1999-2015.

The hourly series of power generation from renewable sources, along with the overall electricity consumption series during November 2014 - October 2015 (Figure 4) shows that, as expected, the wind and solar PV generation was significantly more variable than the geothermal and hydroelectric generation, though the respective shares of year-round RES generation were not much different (HYDRO generation = 55 TWh; WIND+PV generation = 41 TWh). The latter figures account for 17.5% and 13%, respectively, of the overall electricity consumption in the same period (315 TWh).

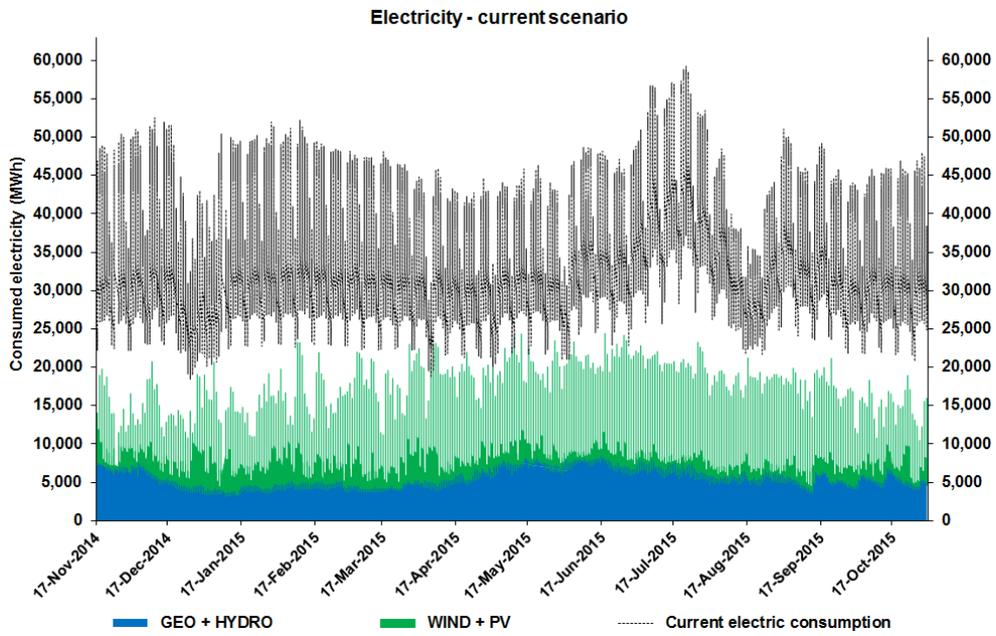

**Figure 4**. Hourly series of RES generation and electricity consumption in Italy, during November 2014 to October 2015.

Back to total energy consumption, Figure 5 shows that, in Italy, electricity accounts for almost half of the TEC.

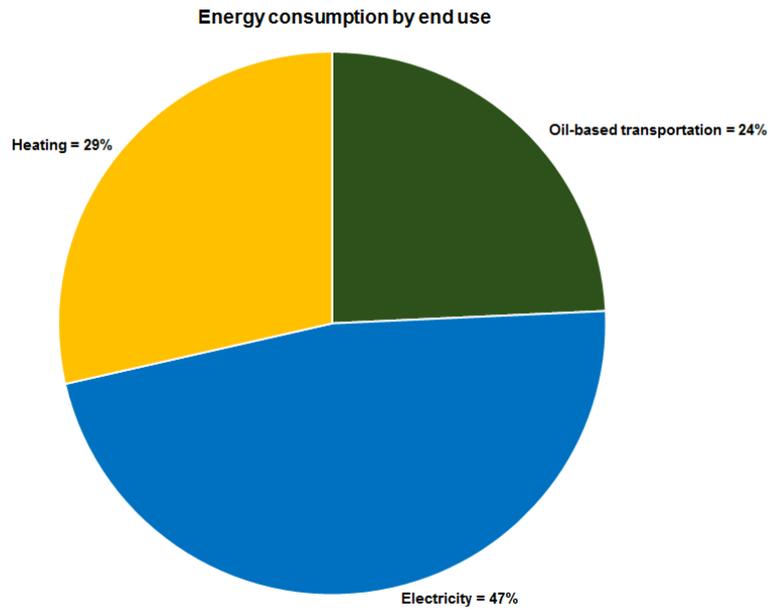

**Figure 5**. Breakdown of total energy consumption into energy end use in Italy, (Nov-2014-Oct-2015). Heating is powered by fossil fuels.

**3 One hundred per cent renewable energy scenario**

The transition to full renewable energy generally requires the electrification of energy end uses,[24] particularly of heating and transport needs, which are currently met by burning fossil fuels.

In a recent work aimed at promoting a rapid transition to 100% clean, renewable energy (The Solutions Project),[25] Jacobson and co-workers assume that Italy, under the renewable-only electricity scenario, would achieve about 34% reduction in end-use power demand due to the much higher efficiency of electric motors over the internal combustion engine; as well as of electric heating with heat pumps and the elimination of energy use for the upstream mining, transport, and/or refining of fossil and biomass fuels. Although the latter is a reasonable hypothesis, guided by the realistic approach mentioned above, in this work we have chosen to assess the power needs deriving from each kind of broad end use – direct use of electricity, heating and transportation – that sum up to provide the overall figure for the whole Italian electricity demand to be covered by RES as such.

To convert the current total energy consumption in Italy into electricity, the average efficiency from fossil fuel heating installations was prudentially assumed equal to as much as 90%. The structure of the distributed electric heating infrastructure was assumed to derive from a 50% heating demand concerning high-temperature heat – requiring electric resistances whose efficiency can be assumed again around 90% - and the other 50% concerning low temperature (<100°C) heat, which can be matched by air-source heat pumps (ASHP). Again, such partition can be regarded as quite conservative, since residential heating is likely to share more than 50% of the overall demand.

ASHPs are assigned an average coefficient of performance (COP) of 4, *i.e.* the ratio between the quantity of heat transferred to the heat sink (useful energy output) and the input electric power, which is typical of relatively mild Italian climate.[26] While ground-source heat pumps generally enjoy higher values of the COP than ASHPs,[27] their higher capital costs and less general applicability lead us to select air-source heat pumps as the reference electric low-temperature heating technology.

On the basis of the above assumptions, the current heating portion of the overall energy consumption (43 MTOE) is first partitioned into high-temperature heating (21.5 MTOE). Due to the assumed efficiency, the latter is simply multiplied by the oil energy content (11.6 kWh/kg), resulting in an electric consumption equal to 251 TWh. The remaining portion of the current heating demand (again 21.5 MTOE), matched by air-source heat pumps, is multiplied by the efficiency of fossil fuel heating installations (90%), then divided by the assumed average COP of the ASHPs (4), and finally multiplied by the oil energy content (11.6 kWh/kg), resulting in additional 57 TWh. Overall, the above provides an additional electric consumption due to heating requirements equaling 308 TWh.

Concerning replacement of oil-based transportation, the current equivalent oil energy content is multiplied by the ratio of the average efficiencies of internal combustion engines and electric vehicles (20% and 80%, respectively),[28] resulting in 107 TWh. Hence, the total electricity consumption including current electricity consumption (315 TWh in 2015) to be matched would amount to 730 TWh, *i.e.* about 2.3 times the current electric demand. Additional demand for heating purposes would share 42%, while additional demand for transportation would follow with 15%.

Unless a country or region has access to suitable energy storage such as large capacity pumped hydroelectric power, balancing and variability are the main problems to address with intermittent renewable energy sources.[29] As mentioned above, a fairly large storage for intermittent wind and solar PV generation is already in place in Italy, based upon the residual capacity of reservoir and basin hydroelectric plants. In order to compute such residual capacity, we assume that that no other storage system is available and that 400 h is the average storage time available for the whole set of the plants under consideration. Thus, we calculate the amount of extra wind and PV energy that can be generated and stored as hydroelectric power in this residual capacity before being conveyed to the grid. Figure 6 shows one of the possible outcomes, with wind capacity around 18,000 MW, *i.e.* about double the current value, and solar PV capacity around 30,000 MW, *i.e.* about 1.6 times higher than the current value.

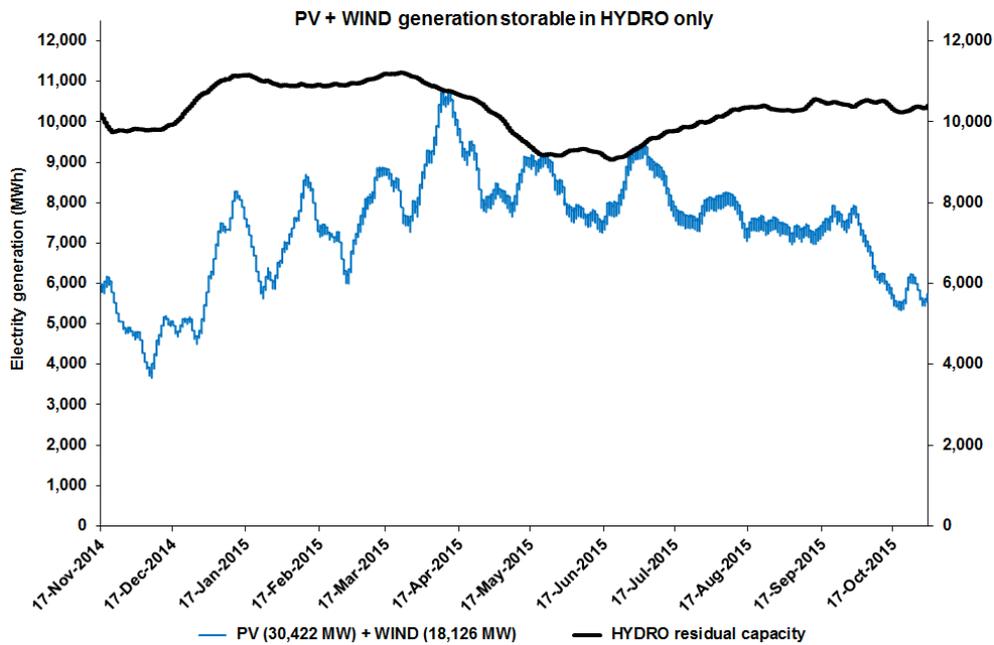

**Figure 6**. Determination of upper bounds for wind and solar PV capacity whose overall generation matches the hydroelectric residual capacity in Italy, during November 2014 to October 2015.

The yearly electricity generation from the wind and solar PV sources combined would then increase to 62 TWh (about 1.5 times over the current output), and its share of current electricity consumption from 13% to about 20%. The sum of the wind and solar PV generation after hydroelectric storage (62 TWh), the current hydroelectric (about 49 TWh) and geothermal (ca. 6 TWh), results in about 117 TWh, *i.e.* 16% of the projected total energy consumption in the 100% electric and renewable scenario.

Although in principle a further increase of the hydroelectric capacity with storage capabilities (reservoir and basin plants) cannot be ruled out, it is unlikely to be sufficient to meet the storage requirements due to the very large additional wind and solar PV generation needed to match the projected total energy (all electricity) consumption.

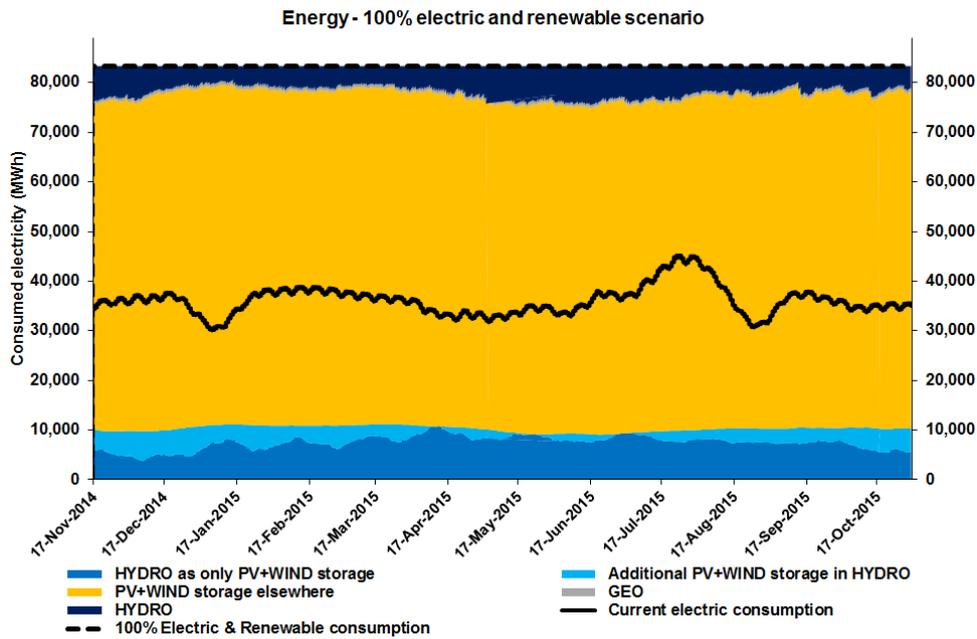

**Figure 7**. Partition of the renewable electric generation matching the overall consumption under the 100% electric and renewable scenario in Italy, simulated during November 2014 to October 2015.

Figure 7 shows the real (observed) electric consumption series and the simulated structure of the hourly renewable electricity generation for the same period under consideration in this study (November 2014 to October 2015) matching the required total electricity consumption. For the sake of simplicity, the annual value of 730 TWh is evenly distributed across 8,760 hours, resulting in an hourly demand equal to about 83,000 MWh. Notwithstanding that at least a fraction of such additional intermittent generation could skip the storage, due to the chance of delivering any excess generation via cross-border interconnections, as in fact it is already occurring,[30] for the purpose of this study it is assumed that the whole WIND+PV generation must be stored.

The hydroelectric generation after the storage of the intermittent PV+WIND generation, is partitioned in its turn into storage shown in Figure 6 and additional storage in hydroelectric basins and reservoirs up to saturation of the respective residual capacity, as shown at the bottom of Figure 7. The chart includes PV+WIND generation after its storage elsewhere (for example, in batteries, stationary or onboard electric vehicles), as well as the geothermal and hydroelectric generation observed during the period. All such generation components match the required electricity consumption every hour.

Under this scenario, the solar PV and wind capacities should be respectively increased up to about 315 GW (16 times more than current capacity) and 190 GW (21 times higher than current capacity), their combined annual generation would reach about 640 TWh (370 TWh from PV, 270 TWh from WIND), out of which more than 550 TWh should be stored elsewhere than in currently available hydroelectric basins and reservoirs, averaging more than 65,000 MWh every hour.

The first outcome of the above analysis concerns the surface requirements (land + building surface) needed for the new solar PV installations as well as the power and number of new wind towers and farms. According to 2002 estimates by the International Energy Agency, with appropriate rooftops and facades covered with commercial solar cells (763,53 km2 of rooftop and 286,32 km2 of facade surface available in 2002, and since then further grown), the potential annual energy production in Italy is 126 TWh.[31] This suggests that only through building integrated photovoltaics, 40% of Italy's current electricity demand, as well as 34% of the respective target generation in the 100% electric and renewable scenario (370 TWh), can be met with country-wide BIPV deployment.

To understand the scope of the transition required to achieve fully distributed and renewable generation, at the end of 2013, when the installed PV power in Italy was

**14**

17,429 MW, out of the 20.3 TWh of solar electricity produced, 80% was fed into the grid and 20% was consumed locally.[32]

Large penetration of renewable energy in Italy too has deeply affected the electricity market, consistently lowering its price and the way pricing is structured.[33] Perez and co-workers have lately shown that achieving very high PV penetration and the displacement of conventional power generation will require an effective electricity remuneration framework in which customers are alerted to take advantage of low-cost electricity when there is a surplus. In turn, this hints to an essential role of the grid operator in balancing the total amount of renewable generation with the rest of the grid.[34] Such distributed generation, furthermore, lowers the load on the grid by generating a significant fraction of power just where it is absorbed, while storage in Li-ion batteries now equipping commercial solar inverters helps reducing the backup energy demand by averaging intermittent power generation in time.

The second and even more important outcome of the present analysis, indeed, concerns the immense capacity needed to store about 550 TWh of additional PV+WIND generation: where all this energy will be stored whereas currently, as put it by Pickard, virtually *none* of the needed bulk storage capacity exists today?[35]

**4 The new role of biomass and solar thermal technologies**

The issue facing policy makers is to identify as soon as possible the efficient and economically viable energy storage technologies on which to invest, in order to have them ready when, at the best in about 50 years or more likely in 10 to 20 years,[36] the supply of fossil fuels will have become scarce and thus most countries will have to run on 100% renewable energy.

Storage, indeed, was the main obstacle identified also by Jacobson and co-workers who recently attempted to solve the related grid reliability problem with 100% penetration of intermittent wind, water, and solar for all purposes in the US, by avoiding expensive chemical battery technology and assuming energy storage of heat in the soil (Underground Thermal Energy Storage) and water, cold in water and ice, and electricity in hydrogen, phase-change materials, and hydropower.[37] Exploring the possibility of having Britain entirely powered by solar PV energy, MacKay, for example, has recently emphasized that 16,000 kg of batteries per person out of 60 million people comprising the UK population would be needed to store the electrical energy consumed throughout the wintertime (battery density = 100 Wh/kg; roughly 100 days of 40 GW average demand, *i.e.* 2356 h×40 GW).[38]

The other major point concerns clean transport technology, which has been hampered for decades by the obsolescence of the battery technology. Things started to change with the advent of state of the art lithium-ion batteries, whose energy density already of about 200 Wh kg$^{-1}$ is improving while cost is falling more rapidly than expected, with prices declining by about 14% per year since 2007 in a standard learning curve effect of 6% to 9% reduction in price for every doubling of production volume.[39] Li-ion batteries built with recyclability in mind will play, we argue herein, a significant role to store solar PV energy in stationary buildings, as it is already happening with hundreds of inverter manufacturers selling inverters equipped with Li-ion battery packs capable to store from a few to several kWh. With cars being produced at 80 million yr$^{-1}$ rate, each needing about 10 kg of lithium, this will simply not be the case as reserves of fossil lithium amount to 13.5 million tons,[40] enough to sustain less than a 17-year supply (13,500,000/800,000 = 16.9).

Put simply, we will not replace fossil, finite resource such as coal, oil and natural gas with another fossil, finite resource; but rather with a renewable energy storage substance such as low cost and abundant renewable hydrogen obtained from evenly distributed biomass in which hydrogen is stored in safe and easily handled polysaccharides with a striking ~5000 Wh/kg energy density.[41]

Indeed, although global efficiency of plant photosynthesis is merely 0.2%, the global primary biomass production is approximately five times the world's energy consumption.[42] Zhang and his teams in the US and in China have advanced scalable cell-free enzyme-based catalysis technology through which the whole hydrogen content of cellulosic biomass is extracted from lignocellulosic agricultural waste.

The technology is rapidly progressing. It will be first commercialized in China in the synthesis of valued bio-based products (vitamin B8 and synthetic starch), and then of amylose to be used as an easily handled solar fuel.[43] While formidable standardisation, industrialisation and upscaling efforts would be needed in front of the immense storage requirements mentioned in this and in related works, this emerging technology appears one of the few viable alternatives.

In order to alleviate the burden represented by the additional electric demand for heating purposes (308 TWh, or 42% of the overall consumption), the so far severely underexploited solar thermal (ST) technology should play a new and far more significant role in Italy's energy future. By the end of 2013, the total installed capacity of glazed collectors (flat plates and evacuated tubes) in Italy amounted to 3,650,000 m$^2$ of collector area, corresponding to 2.555 GW$_{th}$, with around 200 MW$_{th}$ of newly installed solar thermal capacity.[44] With a net thermal energy, deliverable by a standard ST installation (2.13 m$^2$ collector surface area), recently assessed as representative for Italy at the level of 1,407 kWh$_{th}$ per year,[45] the overall thermal generation from the ST installations during the year 2013 can be estimated around 2.4 TWh$_{th}$, or about 940 MWh$_{th}$ per MW$_{th}$ installed capacity. Compared to the European solar thermal benchmark of 264 m$^2$ per 1,000 inhabitants, in Italy the national average as of 2014 was still very low (about 50 m$^2$ per 1,000 inhabitants, with market in 2014 having fallen by 25% compared to the previous year at 268,500 m$^2$ newly installed collector area.[46]

A major shift of current heating technologies is urgently required, with far more solar thermal, geothermal and heat pump technologies needed in the heating sector, is fully justified by the current large share of heating (29%) in final energy demand, as shown in Figure 5. Today's solar thermal technology using evacuated tubes (the most efficient), or flat glazed panels (second best) is perfectly suited to provide low temperature heat (<100°C) not only for hot water delivery but also for space heating, hot water and air conditioning (solar combi+ systems), a condition that in the Mediterranean area has been identified as crucial if solar thermal is to contribute significantly to the long-term heating and cooling demand in the European Union.[47]

Hence, given its very high potential for further deployment in Italy (and elsewhere), a fourfold increase of capacity and low temperature heat generation from the ST technology in order to achieve the European solar thermal benchmark can be reasonably assumed. This would contribute at least additional 7 TWh$_{th}$ that, in turn, converted into electricity demand by heat pumps (COP=4), would alleviate the yearly overall electric burden of the 100% electric and renewable scenario by only about 2 TWh. If, however, coherently with the approach of this study, we assume that the increase of ST capacity parallels the PV one, *i.e.* 16 times the current figure to about 46 GW$_{th}$ (after assuming a 300 MW$_{th}$ capacity increase since end of 2013 to mid-2015), that would contribute about additional 40 TWh$_{th}$ in the form of low temperature heat that, in turn, converted into electricity demand by heat pumps, would alleviate the yearly overall electric burden of the 100% electric and renewable scenario by a more relevant amount of 10 TWh, *i.e.* by 1.4%. While this figure is not so great in the framework of the new scenario, and while it does not change significantly the 550 TWh estimate for the additional PV+HYDRO generation, it would mean about 1 m$^2$ of solar collector area per each inhabitant – a threshold difficult to overcome in the realistic approach adopted in this study.

# 5 Economic analysis

Significant investments in additional renewable capacity comprised of overall 300 GW of PV power and 180 GW of wind power is required, along with key electricity grid and storage investments. The estimated capital costs for this scenario using current full installation costs (€1/W for PV,[48] €1.3/W for wind power[49]) amounts to about €530 billion for energy generation, even though the actual cost will be significantly lower due to rapidly declining prices for both solar modules, wind turbines and Balance of Systems components. Computing from 2017, therefore, the average additional investment needed in renewables until 2050 would be about €16 billion a year (current money). The additional costs for refurbishing and replacing the installations after 2050 are deliberately omitted under the assumption that the savings produced in the electricity bill will compensate such costs.[33]

The eminent role of PV energy, much bigger than wind, is justified by the very low cost of the technology, with 2017 prices of PV modules having already approached the $0.30/W threshold.[50] In addition, we do not differentiate between three different types of PV (residential, commercial and utility), as the latter differentiation is largely academic.

A recent study outlining three energy scenarios aimed at reducing emissions in 2050 by 80% compared to emissions in 1990, concluded that the most technically feasible pathway to decarbonize the Italian energy system would rest on deploying solar and wind technologies, a significant contribution from biomass generation, a moderate but critical role for carbon dioxide sequestration, accompanied by the deployment of more efficient technologies in a number of industrial sectors within the Italian economy, as well as in transport and residential energy uses.[51]

We also ascribe a significant role to biomass, yet in an advanced scenario in which biomass is not merely burned to generate electricity or make first-generation biofuels, but rather used as an abundant source of valued biohydrogen using the multi-enzyme technology first proposed by Zhang and co-workers.[41] Based on recent estimates, the cost of carbohydrate (amylose) produced at industrial maturity could level around $0.3/kg (€0.25/kg at the present exchange rate) or lower.[43] Assuming the specific gravimetric energy content of amylose at about 17 MJ/kg, *i.e.* 4.7 kWh/kg, a rough estimate around €29 billion (current money) can be derived for the cost of storage of 550 TWh of additional PV+WIND generation not storable in hydroelectric basins and reservoirs. Assuming a linear increase of additional PV+WIND capacity during 2017-2050, the overall cost for storage during the same 34-years period would amount to about €500 billion, bringing the overall economic cost of transition towards € 1 trillion.

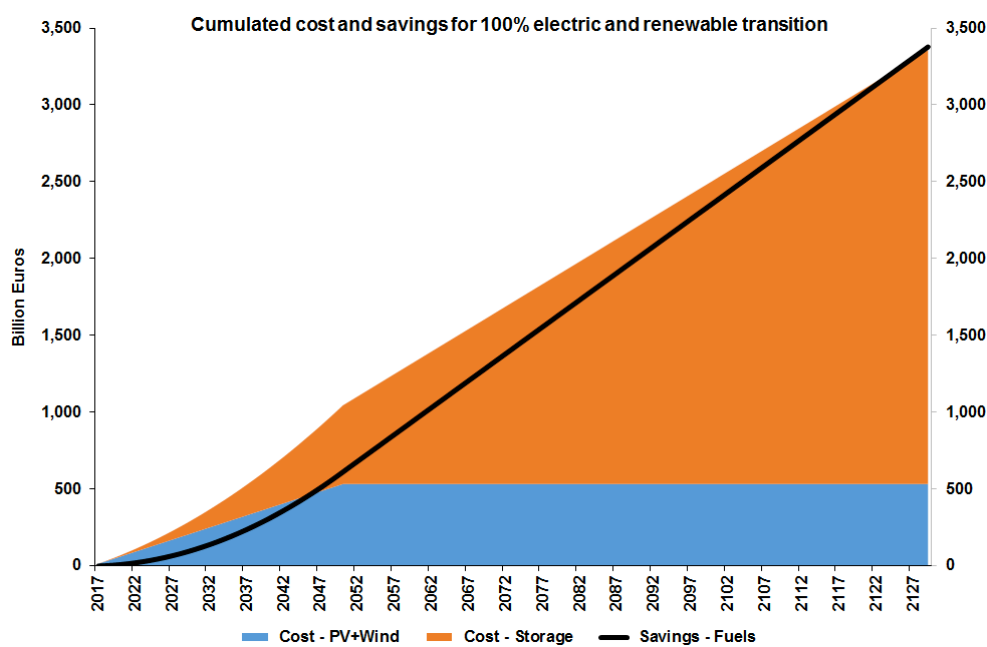

**Figure 8**. Economic balance of the proposed energy transition, partitioned into the cumulated cost of deploying additional PV+WIND capacity, cumulated cost of storage and cumulated savings of fossil fuels.

In the year 2015, the Italian energy bill (oil, natural gas, coal and other solid fuels, net electricity import) was about €35 billion, out of which about €16 billion due to oil import.[52] Assuming constant prices for all fuels, as well as the above-mentioned linear increase of the additional renewable capacity up to the year 2050, the overall cost avoided by the transition would amount to about €600 billion.

Therefore, the overall *net* economic investment needed to perform the transition to 100% electric and renewable energy system in Italy would be limited to approximately €400 billion during the next 34 years, or an average of less than €12 billion per year. Afterwards, the main cost will concern the storage (€29 billion per year, in current money), *i.e.* €6 billion lower than the yearly fuel savings, leading the pay-back time of the investment to occur many decades after 2050, as shown in Figure 8. Although this may appear a long time, it should be considered that the scenario designed in this study will ensure, in the short term, a progressive resilience against both price fluctuations and possible scarcity of fossil fuels, while in the medium term (between 15 and 30 years) it will be the only alternative against a likely severe energy shortage. Moreover, the fuel savings overcome the cost of deploying additional PV+WIND capacity already in 2046, or roughly 30 years from the beginning of the proposed transition: in the meantime, the costs of both the additional renewable capacity and the storage technologies are likely to decrease significantly, leading to a far more attractive economic balance.

**6 Outlook and Conclusions**

We outline a realistic energy transition roadmap for Italy in which the whole energy demand by 2050 is met by electricity generated by low cost renewable sources, mainly solar photovoltaic, wind, and hydroelectric, along with the highly sustainable solar thermal technology to generate low temperature heat. It should be clearly understood that the transition demands to invest for many years to come a significant part of currently available fossil fuels in producing wind turbines, photovoltaic modules, solar collectors, heat pumps, metal wires, concrete, batteries and bioenergy plants to meet the energy demands of modern civilization.[6,15]

Along with the electrification of virtually all energy end-uses and the transition to heat pumps as the standard technology covering the generation of low temperature heat, the analysis identifies large electricity storage requirements that, in our viewpoint, will be met by a technology mix in which biomass will play a pivotal role.

The study deliberately omits to consider grid interconnectedness to balance the generation and demand via interconnection between the Italian and European electricity networks, importing electricity when generation is lower than demand and exporting surplus electricity when generation is greater than demand, as it actually happens in Italy since decades. In *lieu* of it, the key energy balance issue is addressed employing pump storage along with second generation biomass utilization via enzyme-based technology to store any electricity surplus into energy dense carbohydrates fixing the $CO_2$ extracted from the atmosphere.[53] In addition, consumer behavior towards electricity usage will help reduce the amount of balancing required, which in its turn requires reshaping the market structure.[34]

Increasingly, policy makers of forthcoming Italy's Governments elected by people ever more concerned about declining resources, environmental crisis and climate change, will adopt wise energy policies capable to govern the required energy transition towards a fully renewable energy future. The figures and the realistic approach devised in this study will hopefully provide guidance in building such new policies, in which fundamental and industrial research for ever more efficient, reliable and economically viable power generation and storage technologies will play a pivotal role.

## 7 Acknowledgements

This article is dedicated to Professor Vaclav Smil, University of Manitoba, Canada, for all he has done to advance the understanding of energy transitions. Thanks to Professors Yi-Heng Percival Zhang, Virginia Tech and Tianjin Institute of Industrial Biotechnology, Richard Perez, University at Albany, State University of New York, Tomas Kåberger, Chalmers University of Technology, Mark Z. Jacobson, Stanford University, and Derek Abbott, University of Adelaide, for helpful discussion.

**List of abbreviations**

BP = British Petroleum

GEO = Geothermal power

GSE = Italian Manager of Energy Services

MISE = Italian Ministry for Economic Development

MTOE = Million Tons of Oil Equivalent

MJ = Megajoule

MW = Megawatt

MWh = Megawatt hour

PV = Photovoltaic

RES = Renewable Energy Sources

TWh = Terawatt hour

WIND = Wind

# Reference

**1**. Agora Energiewende, Why there was not 100 percent power consumption from renewable energies on Whit Sunday after all, Berlin, 17 May 2016.

**2**. H. Lund, B.V. Mathiesen, Energy system analysis of 100% renewable energy systems—The case of Denmark in years 2030 and 2050, *Energy* **2009**, *34*, 524-531.

**3**. PricewaterhouseCoopers, IIASA, the Potsdam Institute for Climate Impact Research, European Climate Forum, *100% Renewable Electricity-A roadmap to 2050 for Europe and North Africa*, London: 2010.

**4**. Greenpeace, German Aerospace Centre, *The Energy [R]evolution scenario 2015*, Berlin: 2015.

**5**. M. Z. Jacobson, M. A. Delucchi, A Plan to Power 100 Percent of the Planet with Renewables, *Sci. Am.* **2009**, *301*, 58-65.

**6**. V. Smil, *Energy Transitions*, 2nd edition, Praeger Publishers: 2017.

**7**. R. L. Bradley Jr, Renewable Energy: Not Cheap, Not "Green", *Cato Policy Analysis* No. 280, August 27, 1997.

**8**. F. Meneguzzo, R. Ciriminna, L. Albanese, M. Pagliaro, The Great Solar Boom: A Global Perspective into the Far Reaching Impact of an Unexpected Energy Revolution, *Energy Sci. Engineer.* **2015**, *3*, 499-509.

**9**. Global Wind Energy Council, Global Wind Report: Annual Market Update, Brussels, 19 April 2016.

**10**. A. Lee, The global PV price benchmark was nudged down again to $24.2/MWh in an auction by Abu Dhabi for a 350MW solar plant, rechargenews.com, September 20 2016,

**11**. F. Mauthner, W. Weiss, M. Spörk-Dür, *Solar Heat Worldwide*, International Energy Agency, Paris: 2015.

**12**. G. Tverberg, Intermittent Renewables Can't Favorably Transform Grid Electricity, ourfiniteworld.com, 31 August 2016.

**13**. T. Trainer, Critique of the proposal for 100% renewable energy electricity supply in Australia, bravenewclimate.com, June 2, 2014.

**14**. T. E. Hoff, R. Perez, Quantifying PV power Output Variability, *Solar Energy* **2010**, *84*, 1782-1793.

**15**. V. Smil, What I See When I See a Wind Turbine, *IEEE Spectrum*, March 2016, p.27.

**16**. R. Fouquet, Historical energy transitions: Speed, prices and system transformation, *Energy Res. Social Sci.* **2016**, *22*, 7-12.

**17**. BP, *Statistical Review of World Energy 2016*, London: 2016.
Available at the URL: http://www.bp.com/en/global/corporate/energy-economics/statistical-review-of-world-energy/downloads.html.

**18**. GSE, Manager of the Italian Energy Services, Available at the URL: http://www.gse.it/en/Pages/default.aspx. (Accessed: 27th August 2016)

**19**. Terna, *Power consumption and reserve*. Available at the URL: http://www.terna.it/it-it/sistemaelettrico/dispacciamento/stimadelladomandaorariadienergiaedellariservasecondariaeterziaria.aspx. (Accessed: 27th August 2016).

**20**. Terna, *Actual generation of intermittent generation*. Available at: http://www.terna.it/it-it/sistemaelettrico/dispacciamento/previsioneproduzioneeolicadaunit%C3%A0diproduzionerilevanti.aspx. (Accessed: 27th August 2016).

**21**. Ministero dello Sviluppo Economico, *Italian oil consumption data*, Rome: 2016. Available at: http://dgsaie.mise.gov.it/dgerm/consumipetroliferi.asp. (Accessed: 27th August 2016)

**22**. Directive 2009/28/EC of the European Parliament and of the Council, 23 April 2009 available at http://eur-lex.europa.eu/legal-content/EN/TXT/?uri=CELEX:32009L0028

**23**. G. Armani (Terna Rete Italia), Impact of Renewable Generation, *Workshop Greenpeace - Terna "Power30", Rome*, 15 October 2014.

**24**. F. Steinke, P. Wolfrum, C. Hoffmann, Grid vs. storage in a 100% renewable Europe. *Renew. Energy* **2013**, *50*, 826-832.

**25**. M. Z. Jacobson, et al., 100% Clean and Renewable Wind, Water, and Sunlight (WWS) All Sector Energy Roadmaps for 139 Countries of the World, stanford.edu, Draft (version 3), April 24, 2016.

**26**. M. Dongellini, C. Naldi, G. L. Morini, Seasonal performance evaluation of electric air-to-water heat pump systems. *Appl. Therm. Eng.* **2014**, *90*, 1072-1081.

**27**. R. Wu, Energy Efficiency Technologies - Air Source Heat Pump vs. Ground Source Heat Pump, *J. Sustain. Devel.* **2009**, *2*, 14-23.

**28**. L. Albanese, R. Ciriminna, F. Meneguzzo, M. Pagliaro The impact of electric vehicles on the power market. *Energy Sci. Eng.* **2015**, 3, 300–309.

**29**. M. J. Alexander, P. James, N. Richardson, Energy storage against interconnection as a balancing mechanism for a 100% renewable UK electricity grid. *IET Renew. Power Gener.* **2015**, *9*, 131-141.

**30**. F. Meneguzzo, R. Ciriminna, L. Albanese, M. Pagliaro, The remarkable impact of renewable energy generation in Sicily onto electricity price formation in Italy. *Energy Sci. Eng.* **2016**, 4, 194–204.

**31**. International Energy Agency, Potential for Building Integrated Photovoltaics, R*eport IEA-PVPS TZ-4 : 2002*, Paris: 2002.

**32**. Autorità per l'energia elettrica il gas e il sistema idrico, *Monitoraggio dello sviluppo degli impianti di generazione distribuita in Italia per l'anno 2013*, Delibera 14 maggio 2015 225/2015/I/eel, Rome: 2015.

**33**. F. Meneguzzo, F. Zabini, R. Ciriminna, M. Pagliaro, Assessment of the Minimum Value of Photovoltaic Electricity in Italy, *Energy Sci. Engineer.* **2014**, *2*, 94-105.

**34**. R. Perez, K. R. Rábago, M. Trahan, L. Rawlings, B. Norris, T. Hoff, M. Putnam, M. Perez, Achieving very high PV penetration - The need for an effective electricity remuneration framework and a central role for grid operators, *Energy Policy* **2016**, *96*, 27-35.

**35**. W. F. Pickard, Massive Electricity Storage for a Developed Economy of Ten Billion People, *IEEE Access* **2015**, *3*, 1192-1407.

**36**. S. H. Mohr, J. Wang, G. Ellem, J. Ward, D. Giurco, Projection of world fossil fuels by country, *Fuel* **2015**, *141*, 120-135.

**37**. M. Z. Jacobson, A. Delucchi, M. A. Cameron, B. A. Frew, Low-cost solution to the grid reliability problem with 100% penetration of intermittent wind, water, and solar for all purposes, *Proceed. Natl. Acad. Sci.* **2015**, *112*, 15060-15065;

**38**. D. J. C. MacKay, Solar energy in the context of energy use, energy transportation and energy storage, *Phil. Trans. R. Soc A*. **2013**, 371*:* 20110431.

**39**. B. Nykvist, M. Nilsson, Rapidly falling costs of battery packs for electric vehicles. *Nat. Clim. Change* **2015**, *5*, 329-332.

**40**. United States Geological Survey, Mineral Resources Program, *Lithium*, 2015. See at the URL: http://minerals.usgs.gov/minerals/pubs/commodity/lithium/mcs-2015-lithi.pdf

**41**. Y.-H. P. Zhang, A sweet out-of-the-box solution to the hydrogen economy: is the sugar-powered car science fiction? *Energy Environ. Sci.* **2009**, *2*, 272-282.

**42**. Y.-H. P. Zhang, Next generation biorefineries will solve the food, biofuels, and environmental trilemma in the energy-food-water nexus, *Energy Sci. Engineer.* **2013**, *1*, 27-41.

**43**. Y.-H. P. Zhang, Constructing the Electricity-Carbohydrate-Hydrogen Cycle for a Carbon-Neutral Future, SuNEC 2016, Palermo, 7-8 September 2016.

**44**. International Energy Agency, *Country Report - Italy. Status of Solar Heating/Cooling and Solar Buildings - 2015*, Paris: 2015.

**45**. E. Carnevale, L. Lombardi, L. Zanchi, Life cycle assessment of solar energy systems: Comparison of photovoltaic and water thermal heater at domestic scale. *Energy* **2014**, *77*, 434-446.

**46**. European Solar Thermal Industry Federation, *Solar thermal markets in Europe*, Brussels: 2015.

**47**. W. Weiss, P. Biermayr, *Solar Thermal Potential in Europe*, European Solar Thermal Industry Federation, Brussels: 2009.

**48**. B. Willis, PV cost decreases to ensure strong demand in 2016 and beyond - EnergyTrend, pv-tech.org, 5 January 2016.

**49**. D. Milborrow, Global costs analysis -- the year offshore wind costs fell, windpowermonthly.com, 29 January 2016.

**50**. S. Vorrath, New solar glut could push solar module prices as low as 30c/watt, reneweconomy.com.au, 15 September 2016.

**51**. M. R Virdis et al., Pathways to deep decarbonization in Italy, The full report is available at deepdecarbonization.org

**52**. Unione Petrolifera Italiana, Energy balance 2015, The full report is available at
http://www.unionepetrolifera.it/wp-content/uploads/2015/12/Preconsuntivo-UP-2015-21-12-2015-DEF-3.pdf

**53**. Y.-H. P. Zhang, W.-D. Huang, Constructing the electricity–carbohydrate–hydrogen cycle for a sustainability revolution, *Trends Biotechnol.* **2012**, *30*, 301-306.